\documentstyle[aps,graphicx,multicol]{revtex}

\newcommand{\be}{\begin{equation}}
\newcommand{\ee}{\end{equation}}
\newcommand{\ba}{\begin{eqnarray}}
\newcommand{\ea}{\end{eqnarray}}

\def\diff{\mathop{\rm\mathstrut d\!}\nolimits}
\def\C60{C$_{60}$}
\def\etal{{\it et al}}

\begin{document}

\title{Free energy determination of 
phase coexistence in model \C60: \\ A comprehensive 
Monte Carlo study}

\author{D.~Costa\thanks{Corresponding author, e-mail:
{\tt costa@tritone.unime.it}}, G.~Pellicane, M.~C.~Abramo,
and C.~Caccamo}

\address{Istituto Nazionale per la Fisica della Materia (INFM) and 
Dipartimento di Fisica \\Universit\`a di Messina,
Contrada Papardo, C.P. 50, 98166 Messina -- Italy}

\maketitle

\begin{abstract}

  The free energy of the solid and fluid phases of the
  Girifalco \C60 model are determined through 
  extensive Monte Carlo simulations. 
  In this model the molecules interact through a spherical pair potential,
  characterized by a narrow and attractive well,
  adjacent to a harshly repulsive core. 
  We have used the Widom test particle method 
  and a mapping from an Einstein crystal, in order to estimate
  the absolute free energy in the fluid and solid phases, respectively;
  we have then determined the free energy along several isotherms, and the
  whole phase diagram, by means of
  standard thermodynamic integrations.
  The dependence of the simulation's results on the size of the sample
  is also monitored in a number of cases.
  
  We highlight how the interplay between
  the liquid-vapor and the liquid-solid coexistence conditions
  determines the existence of a narrow liquid pocket in the phase diagram,
  whose stability is assessed and confirmed in agreement with previous studies. 
  In particular, the critical temperature follows closely
  an extended corresponding-states rule recently outlined 
  by Noro and Frenkel
  [J. Chem. Phys. {\bf 113}, 2941 (2000)].

  We discuss the emerging ``energetic'' properties of the system, which 
  drive the phase behavior in systems interacting through
  short-range forces [A.~A.~Louis, 
  Phil.~Trans.~R.~Soc.~A~{\bf 359},~939~(2001)], in order to
  explain the discrepancy between the predictions of several structural
  indicators and the results of full free energy calculations,
  to locate the fluid phase boundaries. 

  More generally, we aim to provide extended reference data 
  for calculations of the free energy of the \C60 fullerite
  in the low temperature regime, as 
  for the determination of the phase diagram
  of higher order C$_{n>60}$ fullerenes and 
  other fullerene-related materials,
  whose description is based on the same model adopted in this work. \\

\end{abstract}

\begin{multicols}{2}
\columnseprule 0pt

\section{Introduction}

   We report  an extensive investigation
of the free energy characteristics  
of the Girifalco model of \C60 fullerene~\cite{Girifalco}.
As is well known, this representation hinges on the fact that
\C60  molecules have almost spherical shape and freely rotate 
at sufficiently high temperatures~\cite{Spaepen}. Under these conditions,
the hollow molecular cages  can be assimilated
to spheres whose surface consists of a uniform distribution of carbon 
sites. The overall interaction between two fullerene particles
is then obtained by an integral of the interaction between pairs of  
sites on different cages, eventually yielding an analytic central two-body 
potential. The latter
is  characterized by a harshly repulsive core at short
distance, followed by a deep attractive well which rapidly
decays with the interparticle distance ~\cite{Girifalco}.

The Girifalco model constitutes a prototype system in 
several respects; 
recent studies suggest that 
a similar ``smeared out'' spherical description  
can be attempted for C$_{n>60}$ systems, although
fullerene molecules with $n>60$ can have a sensibly 
non-spherical shape; the cases 
$n=70$, 76 and 84 have in particular been examined 
(see~\cite{Abramo1} and references therein).
On the other hand, more refined calculations of 
the fullerene-fullerene interaction 
yield results very close to those predicted
through the Girifalco model~\cite{Broughton,Pacheco}, and
a similar representation has been used  for
the description of other hollow nanoparticles as carbon onions,
or metal dichalcogenides (also termed inorganic fullerenes)
as GaAs and CdSe~\cite{Safran}.
Moreover, a modification of the Girifalco model,
suitable for the description of solid \C60 at low temperatures, 
has been recently proposed~\cite{Hasegawa0}; this
development seems of particular interest since
fullerites, doped with
organic molecules and upon the injection of electron (or holes),
exhibit a superconducting behavior up to $T=112$~K~\cite{Schon};
an accurate description of such a simple model might
prove useful  for further studies
on the lattice behavior upon impurity doping. 

In our opinion, 
such a possible reference role of the Girifalco model 
for further studies on fullerenes and other systems,
calls for a complete and confident
determination of its phase diagram.
With this purpose in mind,
we have investigated the free energy characteristics
of the \C60 model 
for both the solid and the fluid phase
through extensive Monte Carlo simulations, 
spanning the whole high-temperature region of the phase diagram.

A second general aspect 
of the Girifalco potential 
is related to its short-range nature.
It is useful to recall in this respect that a ubiquitous definition 
of the range of interaction has been recently proposed in Ref.~\cite{Noro}.
As several studies have pointed out,
the most apparent consequence of a reduced interaction length
is the metastability of the liquid-vapor equilibrium,
which is preempted by the fluid-solid coexistence~\cite{short}.
It is actually known that a stable liquid-vapor
coexistence still survives for the model envisaged here,
albeit restricted to a few tens degree temperature 
range~\cite{Cheng,Hasegawa}.
A tiny liquid pocket has been predicted also
in the phase diagram of similar models of 
higher order fullerenes~\cite{Abramo1},
although it is argued
that even a modest reduction of the
range of the forces might cause the disappearance of the liquid phase.
The emerging borderline nature of the Girifalco model implies 
that the overall appearance of its phase diagram
sensitively depends on the details of the interaction potential;
in fact, the initial controversy around the existence
of a stable liquid phase for this system~\cite{Cheng,Mooij} 
has been solved
by taking into account on one side the full role of the attractive
part of the interaction~\cite{Hasegawa2}, and on the other,
a conveniently large simulation sample~\cite{Fucile}.
More generally, it has been recognized in Ref.~\cite{Louis}
that the physical behavior of a wide class of systems 
characterized by short-range interactions, and in particular the onset of
freezing, are substantially affected by the ``perturbative'' part
(with respect to the repulsive core)
of the potential,
rather than being dominated  by excluded-volume and packing
(i.e. by entropic) effects, 
as is the case in the currently accepted {\it van der Waals 
picture} of simple liquids~\cite{Chandler}.  
We here investigate, as a further key purpose of this work,
the same issue for the \C60 model
and try to reconcile the predictions coming from
different structural indicators --- usually 
related to the freezing threshold of several
simple liquids~\cite{Verlet,deltas}
and adopted in 
early calculations for this model~\cite{Cheng,CMHNC,Coppolino} ---
with the results of 
full free energy calculations.

The paper is organized as follows:
we present in Section~II
the model and the simulation strategies;
the results are reported and discussed in Sect.~III,
while Sect.~IV is devoted to the conclusions and 
future perspectives  of our investigation.
A preliminary account of this work 
has been presented elsewhere~\cite{Merida}.

\section{Model and simulation strategies}

The Girifalco potential between two \C60 molecules, 
is written as~\cite{Girifalco}:
\ba\label{eq:pot}
v(r) & = &   -\alpha_1 \left[ \frac{1}{s(s-1)^3} +
                   \frac{1}{s(s+1)^3} - \frac{2}{s^4} \right] \nonumber\\
     & & \quad+\alpha_2    \left[ \frac{1}{s(s-1)^9} +
                   \frac{1}{s(s+1)^9} - \frac{2}{s^{10}} \right] \,,
\ea
where $s=r/d$, $\alpha_1=N^2A/12d^6$, and
$\alpha_2=N^2B/90d^{12}$;
$N=60$ and $d=0.71$~nm are the number of carbon atoms and the diameter,
respectively, of the spherical particles;
$A=32\times10^{-60}$~erg\,cm$^6$ and
$B=55.77\times10^{-105}$~erg\,cm$^{12}$ are constants entering the
Lennard-Jones 12-6 potential
through which two carbon sites on
different molecules are assumed to interact.
The finite distance at which 
the potential~(\ref{eq:pot}) crosses zero and 
the minimum of the potential well depth 
are $\sigma\simeq 0.959$~nm and  
$\varepsilon\simeq 0.444\times10^{-12}$~erg
at $r_{\rm min}=1.005$~nm, respectively.

In order to investigate the coexistence properties
of model~(\ref{eq:pot}), it is required the knowledge of the 
free energies of both the solid and the fluid phase.
As a general strategy, the free energy 
of the system is first evaluated 
all along a supercritical isotherm at temperature $\overline T$
by integrating the pressure $P$ as a function
of the density $\rho$
according to the formula~\cite{HM}:
\be\label{eq:fvsrho}
\frac{\beta F(\rho,\overline T)}{N} = 
\frac{\beta F(\overline\rho,\overline T)}{N} +
\int_{\overline\rho}^\rho \frac{\beta P(\rho')}{\rho'}
\frac{\diff \rho'}{\rho'} \,;
\ee
here $F/N$ is the Helmholtz free energy per particle,
$\beta=1/k_BT$ is
the inverse temperature, 
$k_B$ is the Boltzmann constant and  
$(\overline\rho,\overline T)$ is a thermodynamic state
where the free energy 
is known (see below). 
The free energy at different temperatures 
is then calculated along isochoric paths as:
\be\label{eq:fvst}
\frac{\beta F(\rho,T)}{N} = \frac{\beta F(\rho,\overline T)}{N} 
- \int_{\overline T}^T \frac{U(T')}{Nk_BT'}
\frac{\diff T'}{T'} \,,
\ee
where $U/N$ is the internal energy per particle
of the system.
The equilibrium conditions finally derive from the equality of
the chemical potential and the pressure
of the different phases.

All quantities
entering Eqs.~(\ref{eq:fvsrho}) and~(\ref{eq:fvst})
have been determined 
through standard Monte Carlo simulations at constant volume 
or pressure.
Simulations 
have been mostly carried out
on a sample composed of $N=864$
\C60 particles  enclosed in a cubic box with
periodic boundary conditions. The \C60 interaction
has been considered up to half the box length.

As far as
the determination of the
reference free energy 
$F(\overline\rho,\overline T)$ 
in Eq.~(\ref{eq:fvsrho}) 
is concerned, we have used for the solid phase 
the Einstein crystal method described by
Frenkel and Ladd~\cite{Ladd,Smit}.
Namely,
the \C60 interaction in the
solid fullerite is smoothly transformed,
through a coupling parameter $\lambda$,   
into a corresponding harmonic potential
so that
the configurational energy $U(\{{\mathbf r}\})$ 
takes the following form~\cite{Smit}:
\ba\label{eq:ec}
U(\{{\mathbf r}\}) &=&
U_{\rm C_{60}}(\{{\mathbf r}_0\})+(1-\lambda)
\left[U_{\rm C_{60}}(\{{\mathbf r}\})-U_{\rm C_{60}}(\{{\mathbf r}_0\})\right]
\nonumber \\[4pt]
& & \quad + \lambda \sum_{i=1}^N \alpha ({\mathbf r}_i-{\mathbf r}_{0,i})^2 \,,
\ea
where ${\mathbf r}_{0,i}$ is the lattice position of atom $i$
and $U_{\rm C_{60}}(\{{\mathbf r}_0\})$ is the static contribution 
to the potential energy;
$\alpha$ is the spring constant
of the Einstein crystal.
The free energy difference can then be written as:
\ba\label{eq:einint}
\frac{\beta F(\overline \rho, \overline T)}{N} &\equiv &
\frac{\beta F_{\rm C_{60}}}{N} \nonumber\\
& = & \frac{\beta F_{\rm ein}}{N} + \int_1^0 \diff \lambda \left\langle
\frac{\beta}{N}\frac{\partial U(\lambda)}{\partial \lambda}
\right\rangle_\lambda \,,
\ea
where the configurational free energy of the Einstein crystal is:
\ba\label{eq:fein}
\frac{\beta F_{\rm ein}}{N} &=&
\frac{\beta U_{\rm C_{60}}(\{{\mathbf r}_0\})}{N}
-\frac{3(N-1)}{2N}\ln\left(\frac{\pi}{\alpha\beta}\right) \nonumber \\[4pt]
& & +\frac{3}{2N}\ln N -\frac{\ln V}{N} \,.
\ea
In the equation above,
the last two terms account for the fixed center of mass
constraint at which simulations are performed~\cite{Smit}.

As for the absolute free energy 
of the fluid phase,
the chemical potential $\mu$
has been estimated 
at several intermediate densities
through the Widom test particle
method~\cite{Smit,Widom}.

In parallel with the free energy, the entropy 
of the fluid phase is also systematically analyzed,
given that
in several earlier papers  the onset of freezing
of the Girifalco model has been associated
with the vanishing of the {\it residual multiparticle entropy} $\Delta s$,
namely
\be\label{eq:ds}
\Delta s \equiv s_{\rm ex} - s_2 = 0 \,,
\ee
according to
the one-phase freezing criterion originally proposed in Ref.~\cite{deltas}.
In Eq.~(\ref{eq:ds}),
$s_{\rm ex}$ is the the excess entropy per particle
of the system (in $k_B$ units) and
$s_2$ is defined in terms of the radial distribution function $g(r)$ 
of the system as~\cite{ng}:
\be\label{eq:s2}
s_2 = -\frac{\rho}{2} \int\{g(r)\ln [g(r)]-g(r)+1\}\diff{\mathbf r} \,.
\ee
In previous works
$\Delta s$ has been evaluated using liquid state 
integral equation theories.
Here we report Monte Carlo data for the 
structural and thermodynamics quantities 
entering Eqs.~(\ref{eq:ds}) and~(\ref{eq:s2}), in order to perform
a rigorous test of the theoretical predictions based on 
the criterion~(\ref{eq:ds}).

\section{Results and discussion}

\begin{figure}
\begin{center}
\includegraphics[width=10cm,angle=-90]{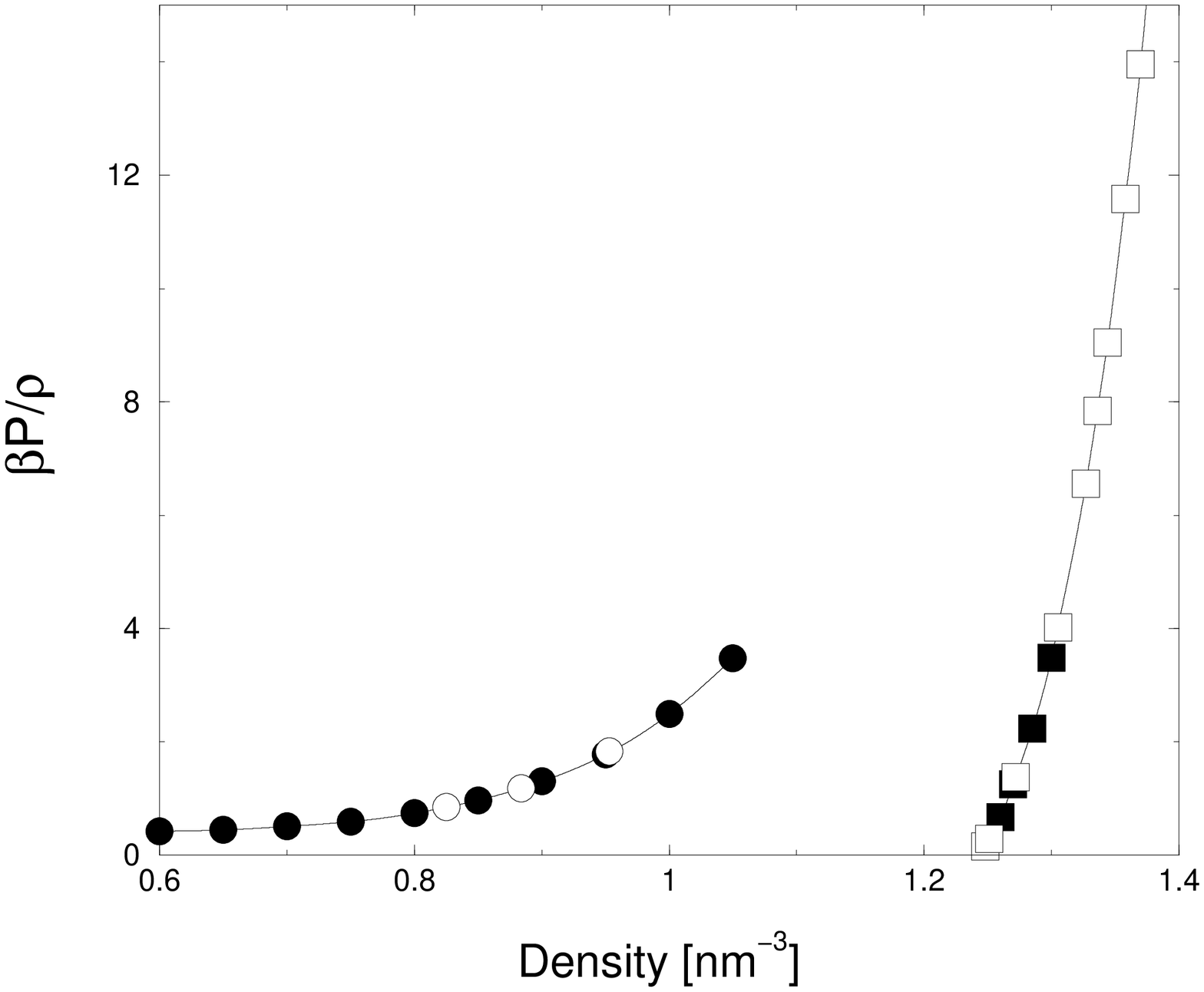}
\bigskip
\caption{Equation of state 
of the Girifalco model
in the fluid (circles) and solid (squares) phases
at $T=2100$~K, as obtained through NVT (solid symbols)
and NPT (open symbols) Monte Carlo simulations.
}\label{fig:bpr}
\end{center}
\end{figure}

The solid and liquid branches
of the equation of state,
to be integrated in~Eq.~(\ref{eq:fvsrho}),
have been calculated 
at the supercritical isotherm
$\overline T=2100$~K through both NVT and NPT Monte Carlo simulations,
in order to
crosscheck the predictions of the two algorithms.
Five to six runs of 25\,000 steps have been  performed at each
thermodynamic point  investigated.
The compressibility factor $\beta P/\rho$ is shown in Fig~\ref{fig:bpr}.

\begin{figure}
\begin{center}
\includegraphics[width=7cm,angle=-90]{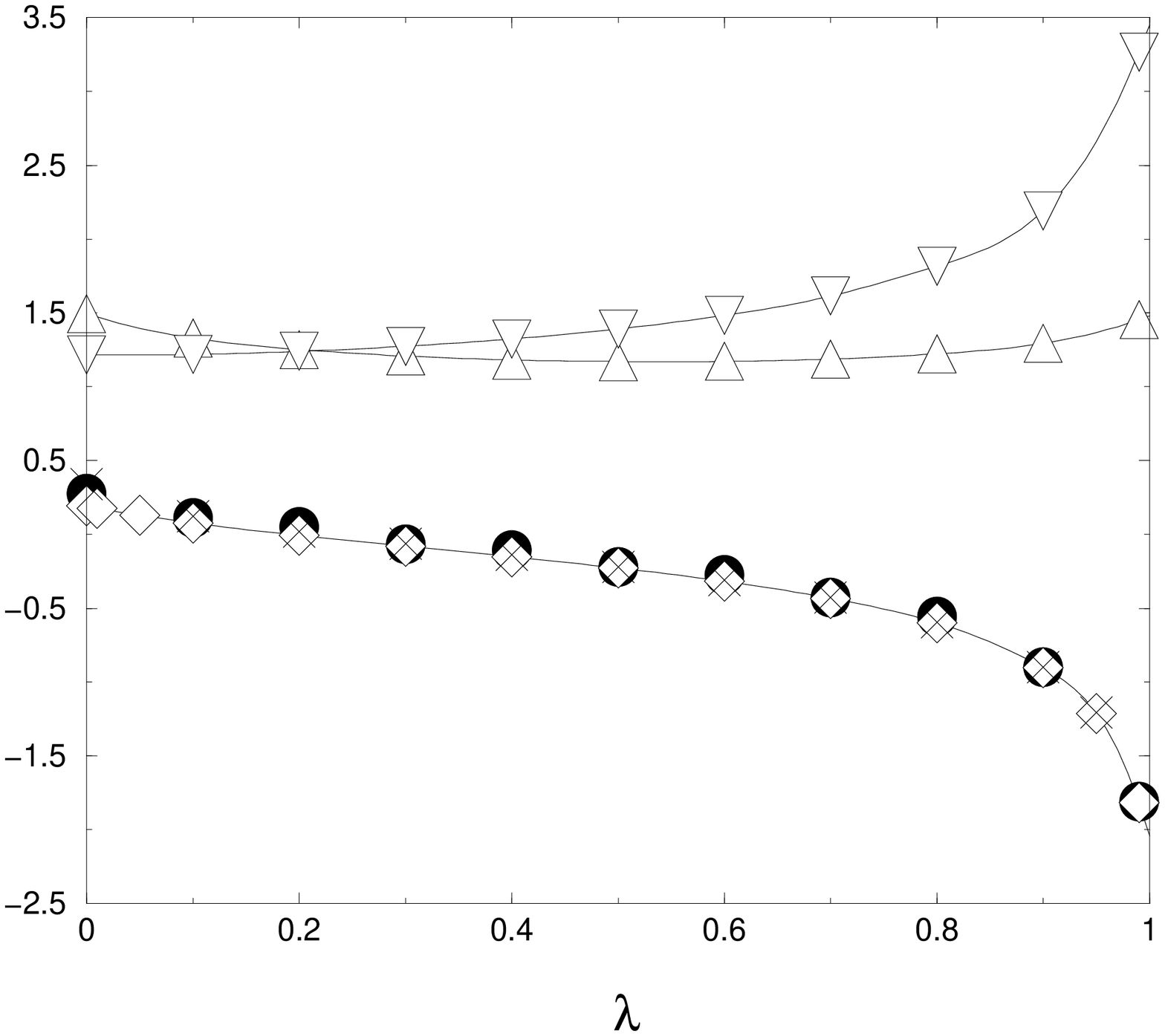}
\includegraphics[width=7cm,angle=-90]{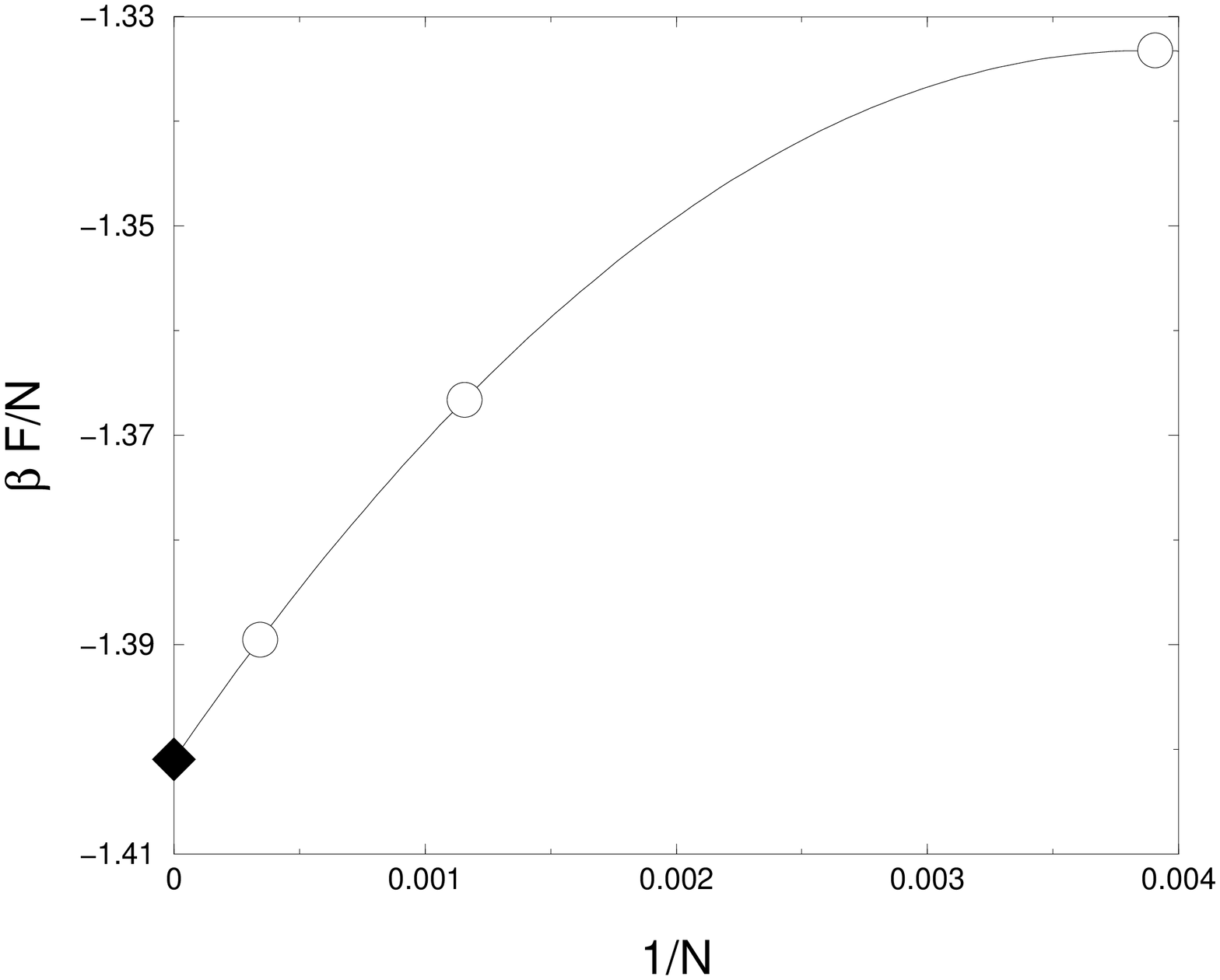}
\bigskip\bigskip
\caption{Results of the Einstein crystal procedure
for the determination of the free energy in the solid phase
at $T=2100$~K and $\rho=1.375$~nm$^{-3}$. 
Left: integrand 
in Eq.~(\protect\ref{eq:einint}) (circles),
resolved into the harmonic (downward triangles) and \C60 (upward triangles)
contributions, see Eq.~(\protect\ref{eq:ec}).
Simulation results obtained
with $N=256$ (diamonds) and $N=2916$ (crosses) particles
are also shown.
Lines are smooth interpolations of the data points.
Right: size dependence of the free energy of the \C60 crystal;
the diamond indicates the extrapolation (solid line) to
the thermodynamic limit.
}\label{fig:ein} 
\end{center}
\end{figure}

Results 
for the absolute free energy of the solid phase at $\overline T=2100$~K
and $\overline \rho=1.375$~nm$^{-3}$, as obtained through the Einstein
crystal method, are reported in Fig.~\ref{fig:ein}.
Constant volume simulations have been carried out
for several values of the switching parameter $\lambda$.
The spring constant is set to a value 
$\alpha/\varepsilon=490$, 
which makes the interactions in 
the pure Einstein crystal as close as possible 
to those of the original system, so to optmize 
the accuracy of the numerical integration 
scheme of Eq.~(\ref{eq:einint})~\cite{Smit}.
We have analyzed
the free energy dependence on the system size;
results with 256 and 2916 particles 
are shown in Fig.~\ref{fig:ein}.
It appears that
the effect of $N$ on the estimate of the free energy is small but
systematic; in the thermodynamic limit ($N \to \infty$) 
the smooth extrapolation
in the bottom panel of Fig.~\ref{fig:ein} 
yields $\beta F/N (\overline \rho=1.375~{\rm nm}^{-3}, 
\overline T=2100~{\rm K})\simeq-1.401$.

\begin{figure}
\begin{center}
\includegraphics[width=7cm,angle=-90]{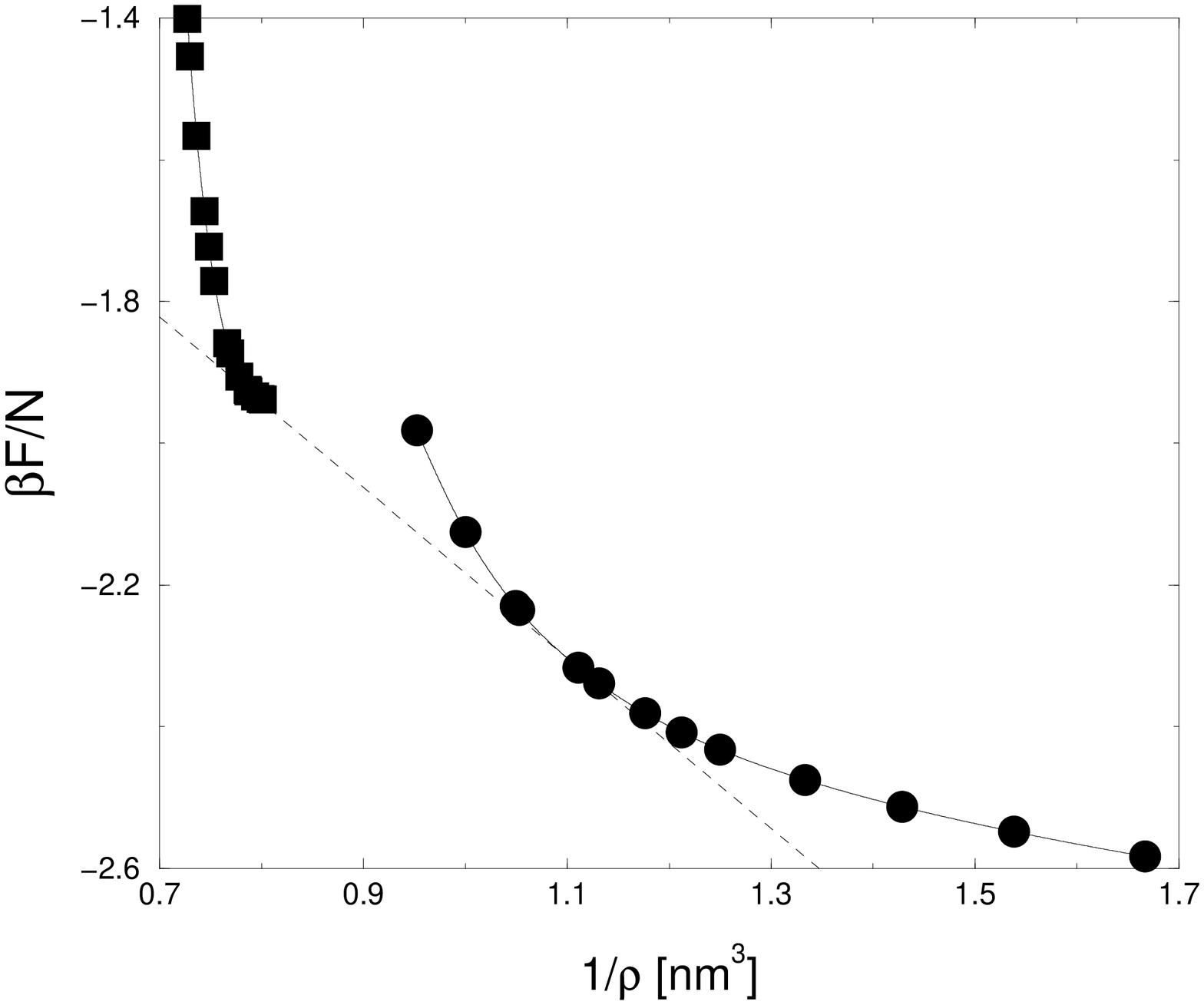}
\includegraphics[width=7cm,angle=-90]{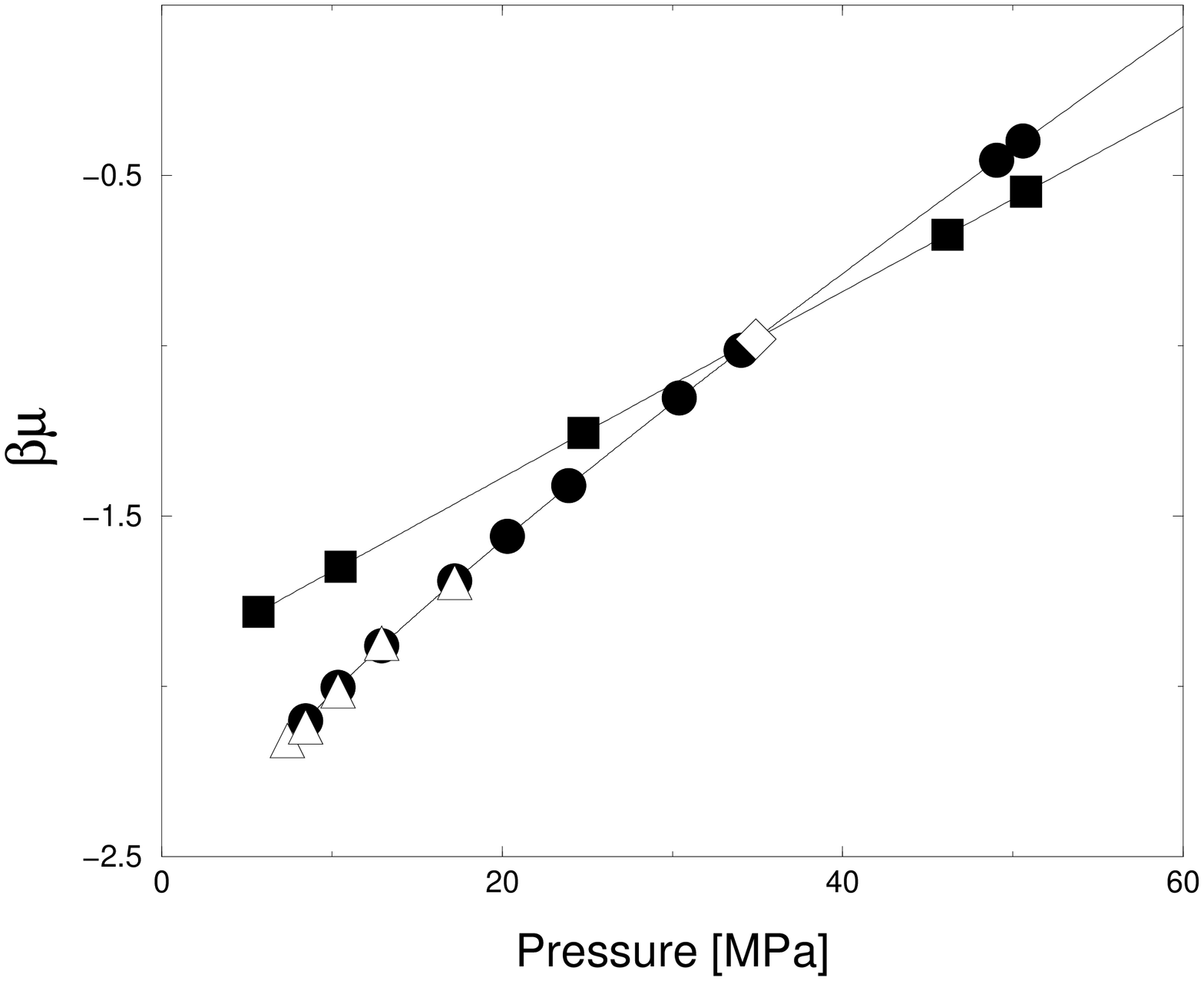}
\bigskip\bigskip
\caption{Free energy (left) and chemical potential (right)
of the \C60 model 
along the isotherm $T=2100$~K. Squares: solid phase; circles:
fluid phase. Solid lines are smooth interpolations of the data points.
In the left panel the common tangent construction (dashed line)
is shown. In the right panel the diamond locates the coexistence
conditions; 
the direct estimates of the chemical 
potential (triangles), based on the Widom 
test particle method, are also reported. 
}\label{fig:fref} 
\end{center}
\end{figure}

As for the fluid phase,
the excess
chemical potential at $\overline T=2100$~K
and $\overline \rho=0.60$~nm$^{-3}$ has been calculated
through the Widom test particle method as $\beta\mu_{\rm ex}=-1.523$.
 Several tests at higher densities have also been conducted,
in order to assess the results obtained via
Eq.~(\ref{eq:fvsrho}).
The free energy at $T=2100$~K is  
shown in the top panel of Fig.~\ref{fig:fref}, along with
the common tangent construction, which determines
the coexistence conditions. 
The $\mu$ {\it vs} $P$ behavior is displayed
in the bottom panel of Fig.~~\ref{fig:fref}; it emerges that
the thermodynamic 
integration of Eq.~(\ref{eq:fvsrho})
fully agrees with the direct estimate of the chemical potential
based on the Widom technique.

\begin{figure}
\begin{center}
\includegraphics[width=10cm,angle=-90]{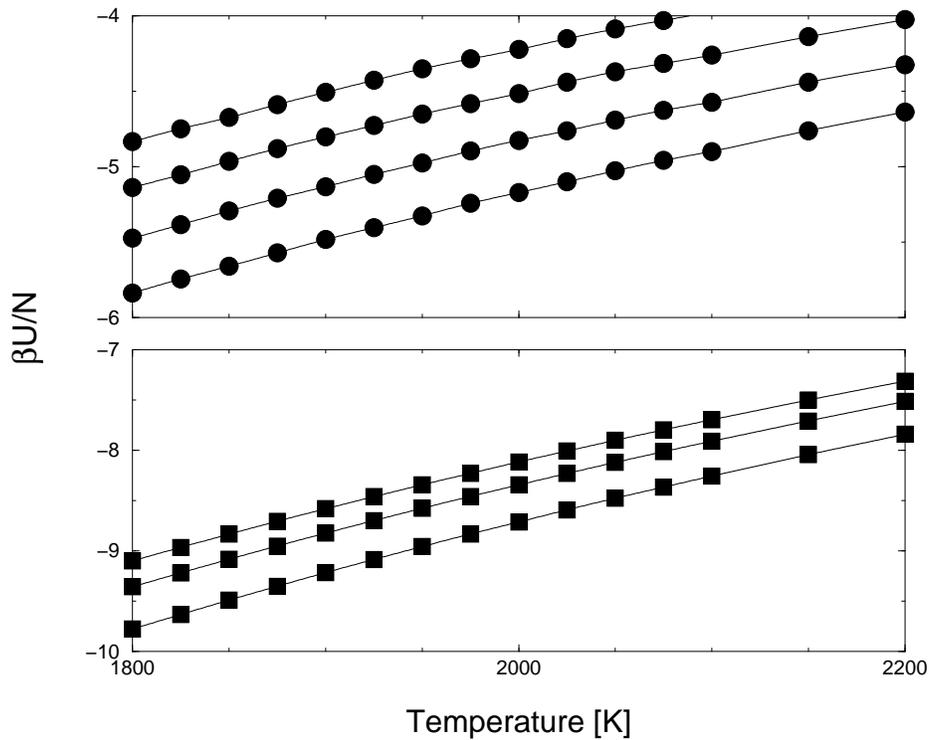}
\bigskip
\caption{Internal energy per particle in the fluid (top)
and solid (bottom) phases. 
The behavior along
the isochores
$\rho=0.80$, 0.85, 0.90, and 0.95~nm$^{-3}$
(top panel, from top to bottom) and
$\rho=1.251$, 1.27, and 1.305~nm$^{-3}$
(bottom panel, from top to bottom) is shown.
Lines are smooth interpolations of the data points.
}\label{fig:bu}
\end{center}
\end{figure}

\begin{figure}
\begin{center}
\includegraphics[width=9cm,angle=-90]{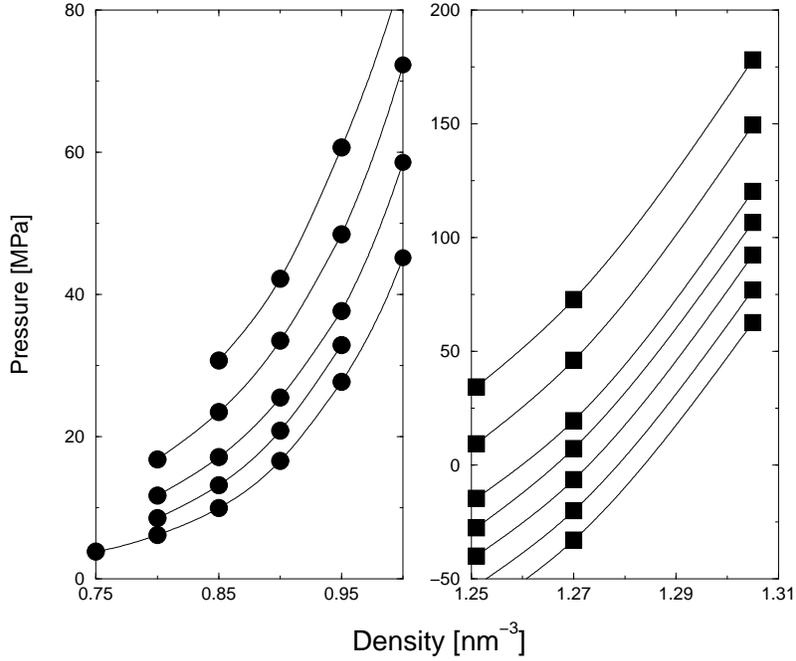}
\bigskip
\caption{Equation of state in the fluid (left) 
and solid (right) phases.
The behavior along the isotherms
$T=2200$, 2100, 2000, 1950, and 1900~K
(left panel, from top to bottom) and
$T=2200$, 2100, 2000, 1950, 1900, 1850, and 1800~K
(right panel, from top to bottom) is shown.
Lines are smooth interpolations of the data points.
}\label{fig:press}
\end{center}
\end{figure}

\begin{figure}
\begin{center}
\includegraphics[width=9cm,angle=-90]{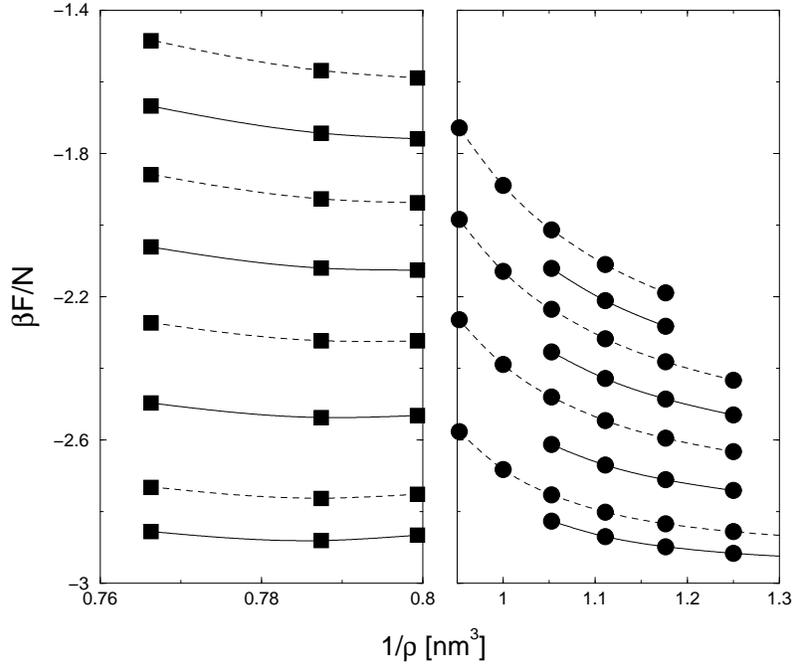}
\bigskip
\caption{Helmholtz free energy per particle in the solid (left)
and fluid (right) phases.
The behavior along the isotherms
$T=2200$, 2150, 2100, 2050, 2000, 1950, 1900, and 1875~K
(from top to bottom) is shown.
Solid lines are guides to the eye; dashed lines
are obtained by integrating Eq.~(\ref{eq:fvsrho}), see text.
}\label{fig:helm}
\end{center}
\end{figure}

 Starting from the knowledge of the free energy along the
isotherm $\overline T=2100$~K, the free energy at different temperatures
has been  obtained through Eq.~(\ref{eq:fvst}). 
We have examined the isochores
$\rho=1.25$, 1.27, and  1.30~nm$^{-3}$ in the solid phase and
the density range $\rho=[0.70 - 1.00]$~nm$^{-3}$ with steps 
$\Delta \rho=0.05$~nm$^{-3}$ in the liquid phase,
descending down to $T=1800$~K,
with temperature intervals $\Delta T=25$~K.
Simulations have been  carried out with
$N=864$ particles at constant density;
four to eight cumulation runs of 10\,000 steps 
at each state point are sufficient to yield
accurate internal energies estimates (see Fig.~\ref{fig:bu}), so to allow
a smooth interpolation 
for the integration in Eq.~(\ref{eq:fvst}).
The pressure and the free energy
along several isotherms are shown
in Figs.~\ref{fig:press} and~\ref{fig:helm}, respectively;
fully consistent results are obtained 
if we determine the free
energy along 
an isochoric path first, and then integrating Eq.~(\ref{eq:fvsrho})
at constant temperature. The accuracy of this global check is also 
evidenced in Fig.~\ref{fig:helm}.

In order to analyze 
the thermodynamic properties 
of the system in the low density regime $\rho=[0.05 - 0.20]$~nm$^{-3}$,
we have calculated  the chemical potential of the vapor phase
along the isotherm $T=1900$~K
through the Widom technique, 
and then
estimated through Eq.~(\ref{eq:fvst}) 
the free energy in the temperature range $T=[1800 - 1900]$~K.

\begin{figure}
\begin{center}
\includegraphics[width=8.5cm,angle=-90]{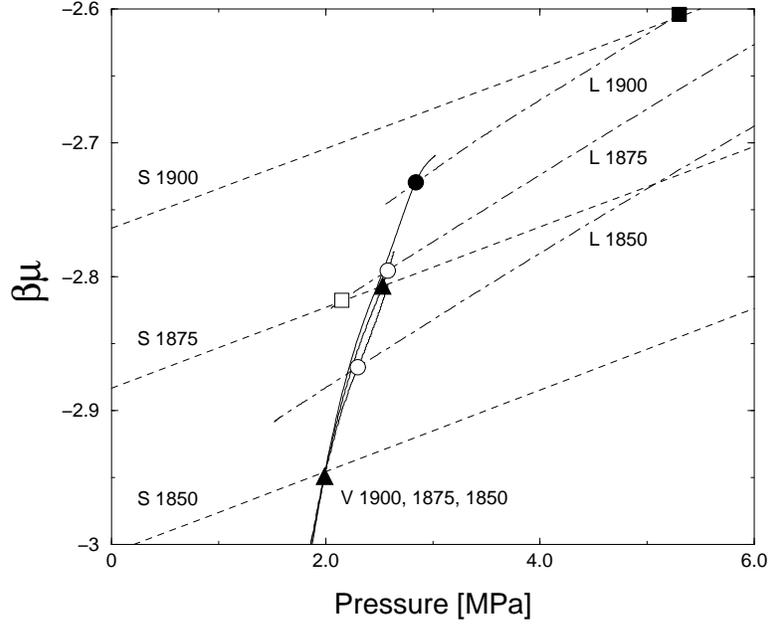}
\bigskip
\caption{Chemical potential {\it vs} pressure
in proximity of the triple point, 
in the vapor (V, full lines), 
liquid (L, dot-dashed lines) and solid (S, dashed lines) phases 
at $T=1900$, 1875, and 1850~K.
Liquid-solid (squares), liquid-vapor (circles) and
solid-vapor (triangles) coexistence points are shown;
full and open symbols refer to stable and 
metastable equilibria, respectively. 
}\label{fig:mu}
\end{center}
\end{figure}

As is visible in Fig.~\ref{fig:mu}, where
the chemical potential of the different phases is shown,
liquid-solid equilibrium is stable at $T=1900$, while 
a solid-vapor coexistence takes place at $T=1850$~K.
The intermediate temperature $T=1875$~K is characterized by 
almost a comparable value of the chemical potential of
the various phases. We thus estimate
that the triple point temperature is 
$T_{\rm tr} \simeq 1875$~K, at the pressure
$P_{\rm tr} \simeq 2.4$~MPa; 
we then obtain from the equation of state
$\rho_{\rm tr} \simeq 0.74$~nm$^{-3}$, in close agreement 
with the free energy results of Ref.~\cite{Hasegawa},
$T=1880$~K and $\rho=0.74$~nm$^{-3}$.
The features in the $\mu$ {\it vs} $P$ behavior which 
determine the narrow temperature width 
of the liquid pocket are clearly illustrated in Fig.~\ref{fig:mu}, 
where it emerges that the 
solid and the liquid free energy branches 
have dissimilar slopes and considerably different spacing
under equal variations of temperature.
These two circumstances cause a fairly rapid shift of the 
liquid-solid intersection points and hence of the 
corresponding equilibrium parameters.
Conversely, the vapor 
branch is hardly sensitive to temperature variations,
and already at $T \gtrsim 1900$~K 
tends to loose any further intersection
with both the solid and the liquid branch.
The only surviving liquid-solid equilibrium therefore 
fully characterizes the system behavior
for $T > 1950$~K. 

\begin{figure}
\begin{center}
\includegraphics[width=8.5cm,angle=-90]{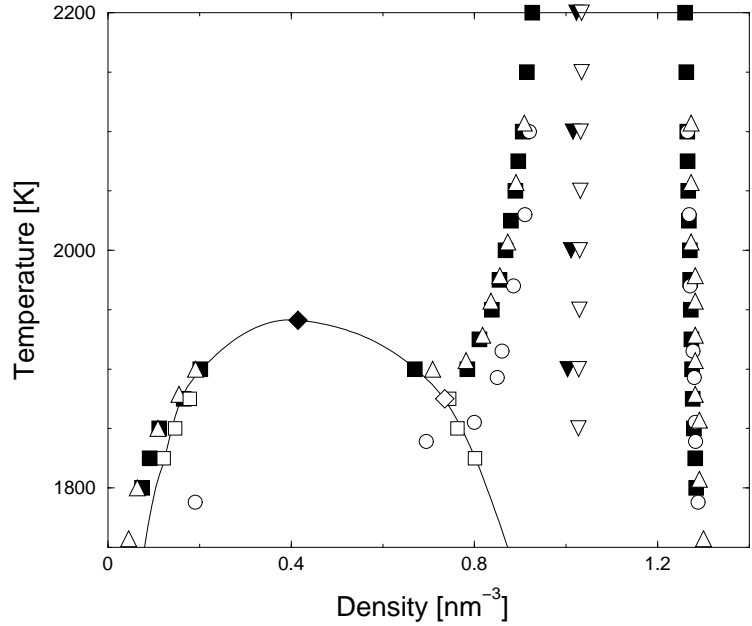}
\bigskip
\caption{Phase diagram of the Girifalco model according
to the free energy investigation of this work (solid squares,
coexistence points; open squares, metastable liquid-vapor
separation).
The line represents the Gibbs Ensemble Monte Carlo predictions
for the liquid-vapor coexistence~\protect\cite{Fucile}.
The critical (full diamond) and triple (open diamond)
points are also shown.
Open upward triangles and circles are the 
simulation results 
of Ref.~\protect\cite{Hasegawa} and~\protect\cite{Mooij}, respectively.
Open downward triangles: $\Delta s=0$ {\it locus};
solid downward triangles: freezing line
of the hard-sphere fluid corresponding to the \C60 model
through the WCA prescription (see text).}
\label{fig:phdia}
\end{center}
\end{figure}

The phase diagram of the system is displayed 
in Fig.~\ref{fig:phdia}, along with our previous 
Gibbs Ensemble Monte Carlo (GEMC)
determination of the binodal line~\cite{Fucile}.
Distinct liquid-vapor and liquid-solid
equilibria take place at temperatures slightly higher than 1875~K; 
as far as the liquid-vapor coexistence is concerned,
we observe a remarkable agreement between free energy calculations
and GEMC results.
At lower temperatures, the 
binodal points are metastable with respect to
the freezing line;
in this case as well, the GEMC approach fully reproduce
the free energy data. 
We thus retain 
the GEMC results 
$T_{\rm cr} \simeq 1940$~K,
$\rho_{\rm cr} \simeq 0.42$~nm$^{-3}$, and
$P_{\rm cr} \simeq 2.7$~MPa,
obtained in Ref.~\cite{Fucile} with 1500 \C60 particles,
as a reliable estimate 
of the critical point parameters.
We note in this context that 
an extended correponding-state behavior, 
which states a linear relationship between the critical temperature
and the range $R$ of the interaction potential,
has been recently outlined in Ref.~\cite{Noro}. $R$
is defined by a map of the potential into
an effective square well interaction 
with the same second virial coefficient.
We obtain for the \C60 model $R=0.16$, a value immediately over
the minimum threshold of a stable liquid-vapor equilibrium, 
$R=0.13-0.15$~\cite{Noro};  the extended rule's estimate
for the critical temperature is then $T=1922$~K, 
in fair agreement with the above 
$T_{\rm cr}=1940$~K result.

A comparison with 
the phase diagram determined by Hagen and coworkers~\cite{Mooij}
is reported in Fig~\ref{fig:phdia}.  
It has been conjectured, on the basis of a theoretical
density-functional investigation~\cite{Hasegawa2},
that the emerging discrepancy 
between Ref.~\cite{Mooij} results and subsequent calculations 
might be due to an early cutoff of 
the \C60 interactions, which substantially affects
the location of the liquid-vapor binodal line.
We argued on the other hand~\cite{Fucile},
that a fairly large simulation sample should be employed
in this case,
in order to take into account
the peculiar density fluctuations in the GEMC simulations,
which act to destabilize the liquid-vapor separation.
We note that, while the 
overall fluid phase boundaries can sensitively depend on the 
interaction details (and hence on the cutoff),
truncation effects play a minor role
in the determination of the melting line;
the latter appears indeed
coincident with our estimate, and almost independent on 
temperature variations, ranging from $\rho=1.26$~nm$^{-3}$ 
at $T=2200$~K 
to $\rho=1.28$~nm$^{-3}$ at $T=1800$~K.
Our results positively agree 
with the phase diagram obtained in Ref.~\cite{Hasegawa}
(also shown in Fig.~\ref{fig:phdia}),
where a slightly higher critical temperature $T=1954 - 1980$~K
is reported.

\begin{figure}
\begin{center}
\includegraphics[width=10cm,angle=-90]{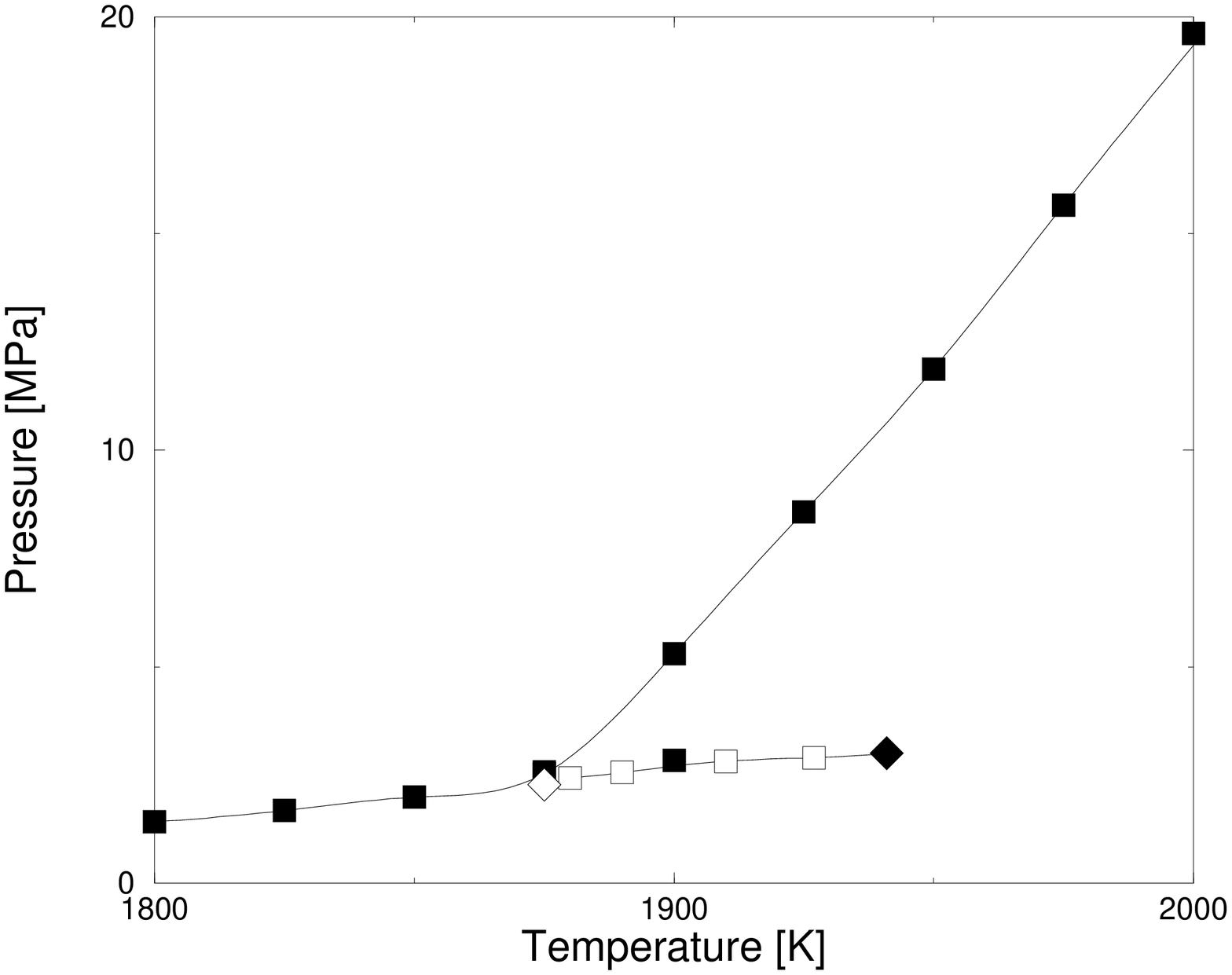}
\bigskip
\caption{Phase diagram 
in the $P-T$ representation. Solid squares, this work;
open squares, Gibbs Ensemble 
liquid-vapor coexistence points~\protect\cite{Fucile}.
The triple (open diamond) and critical (solid diamond) points
are also shown.
Lines are intended as guides to the eye.
}\label{fig:pt}
\end{center}
\end{figure}

The phase diagram in the $P-T$ representation is displayed
in Fig.~\ref{fig:pt}, where it appears that 
also the pressure range of the liquid phase is rather 
restricted, spanning only a few tens bar
over the triple point pressure $P_{\rm tr} = 2.4$~MPa.

We now turn to the indications on the freezing conditions
obtained through
the one-phase criterion expressed by  Eq.~(\ref{eq:ds}).
As is visible in Fig.~\ref{fig:phdia},
the $\Delta s=0$ {\it locus} tends to overestimate the coexisting fluid
density, thus affecting the location of the 
triple point in the phase diagram.
We remark that a similar
trend would emerge if
the Hansen-Verlet prescription~\cite{Verlet,HM} 
for the height of the first peak of the 
structure factor were used~\cite{Abramo1}.
In fact, the close correspondence between the two interpretations
of the freezing transition in terms of structural
indicators,  has been recently 
pointed out in Ref.~\cite{spg}.
Similarly, the behavior of the internal energy, as well as 
the height of the first peak of the radial
distribution function, show a non-monotonic behavior~\cite{Coppolino}, 
suggesting that the system 
becomes unstable against the phase separation 
around the density $\rho\simeq 1.0$~nm$^{-3}$ (see also Ref.~\cite{Cop2}).

 It thus appears that 
 such indicators
 identify a very restricted range 
(if not a unique {\it locus}) of density
 {\it vs} temperature 
 states, over which a structural reorganization of the fluid  phase
 should be tendentially established, in order 
 to satisfy purely excluded-volume, or
 equivalently entropic, demands. The almost vertical disposition
 of the $\Delta s=0$ line (see Fig.~\ref{fig:phdia}), 
which means that 
the density keeps constant along the {\it locus} irrespective
of the temperature, clearly reflects the substantial 
absence of any energy scale associated with such an indication. 
It is known at present that
 the vanishing of the residual multiparticle entropy accurately predicts
 the thermodynamic freezing threshold for a wide class of simple fluids,
including hard-core models and the Lennard-Jones
potential, in both 
three and two dimensions~\cite{spg}.
In these models, 
the interparticle potential is dominated by steric effects, while 
the attractive forces can be treated as 
a perturbation to the inherent hard-sphere system, 
which essentially drives the liquid behavior, an approach
commonly called the {\it van der Waals picture} of fluids~\cite{Chandler}.
This is furthermore illustrated 
in Fig.~\ref{fig:phdia}, where it emerges that the reference
hard-sphere system for the \C60 model, obtained by splitting
the potential into a repulsive and a perturbative part 
in the WCA fashion (after Weeks \etal~\cite{WCA}), 
also freezes around the {\it locus} of vanishing
residual multiparticle entropy, $\rho(T)\simeq 1.0$~nm$^{-3}$.

We argue that
the discrepancy between one-phase indicators and full free energy 
calculations about the freezing transition of the system at issue,
leads to a different scenario,
 where the 
 strong attractive and rapidly decaying well 
in the interaction potential critically
 affects the conditions for the onset of the solid-liquid transition.
 The latter is driven in this case 
 by ``energetic'' rather than by entropic effects, in agreement with
 several indications recently collected
for systems with short-range interactions by A.~A.~Louis~\cite{Louis}. 
 It is shown in Ref.~\cite{Louis} that for this class of fluids
most of the features exhibited
by hard-sphere dominated systems do not arise,
resulting in particular in the anticipation 
of the frezing threshold to
 lower densities than those predicted by solely
 structural conditions. 
In this respect, the lack of an accurate 
estimate of the freezing line of model \C60
upon use of either the residual multiparticle entropy or the Hansen-Verlet
prescriptions, must be interpreted 
less a ``failure'' of the criteria themselves, than
a manifestation of their limited applicability in the present context,
both indicators predicting the freezing threshold of the fluid
on the basis of almost purely entropic requirements. 

It appears in conclusion that
the phase behavior of the Girifalco \C60 model
can be consistently understood in terms 
of general properties of systems interacting
through short-range forces.
Nevertheless, several issues are still open to
further investigations and 
we refer firstly to the physical meaning
to be associated to the fluid phase boundaries 
signalled by the above structural indicators.
We note, on the other hand, the lack of clear indications
of the freezing transition at a thermodynamic level, 
both in the internal energy and
in the pressure, as documented in 
Figs.~\ref{fig:bu} and~\ref{fig:press}. 
More generally, the question  of a detailed description
of the microscopic behavior of the \C60 model --- as
of other fluids interacting through short-range forces --- 
already raised in Ref~\cite{Ashcroft}, is still unsolved. 
The formation of metastable clusters
of strongly correlated particles
in such fluids has been discussed in some recent papers~\cite{Foffi},
on the basis of the short-range
nature of the interaction potential; however, in a recent molecular
dynamics study~\cite{Ballone} of quite a similar model,
we have not been able to identify
any net precursor, at a microscopic level,
either of the fluid-solid threshold,
or of the incipient crystallization of the liquid phase.

\section{Conclusions and perspectives}

 The free energy of the solid, liquid and vapor phases of the Girifalco 
 model of \C60 has been studied by means of extensive 
 Monte Carlo simulations at constant density or pressure;
 the full phase diagram of the system has then been 
reconstructed on such a basis.
 It is confirmed by this comprehensive investigation 
 that a stable liquid phase for this system
 exists, albeit confined to
 a rather restricted temperature range: we confidently estimate
 the triple and critical 
 temperatures as $T_{\rm tr}\simeq 1875$~K 
and $T_{\rm cr}=1940$~K, respectively. 
 The pressure range of the liquid phase 
 also appears rather narrow,
 spanning only a few tens bar
 over the triple point pressure.
Our results illustrate  how
 the interplay of various free energy branches 
determines the overall appearance of the phase diagram, and
in particular the narrow extension of the liquid pocket.

 The estimate of the freezing conditions,
based on several structural indicators, 
 is also critically discussed. 
 It turns out that
 such indicators
 identify a practically unique thermodynamic {\it locus}, 
 where the fluid phase becomes unfavoured 
 due to entropic requirements. This {\it locus} almost
 coincides with the true thermodynamic
 freezing line for systems
 whose phase behavior is dominated by steric effects. 
 For the model at issue, 
 the solid-liquid transition is instead strongly affected by 
 the deep, short-range attractive well in the interaction potential. 
 As a result, the freezing transition  of the fluid is driven
 to lower densities, mainly
 by energetic effects, in agreement with 
 a scenario recently proposed by other authors.

As far as further studies 
are concerned, we are currently assessing,
against the wide set of data produced in this work,
the performances of several 
refined integral equation theories of the liquid state~\cite{Pellican} 
and perturbation approaches, in order to describe 
on a full theoretical ground the phase
diagram and the free energy properties of the Girifalco model.
A refinement of our preliminary report on the phase diagram
of higher order fullerenes C$_{n>60}$~\cite{Abramo1}
is also in progress.
Finally, we plan to investigate the \C60 properties,
as well as the phase behavior
of systems constituted by doped fullerites,
in the low-temperature region of the solid phase.

\end{multicols}

\clearpage


\bigskip

\bigskip

\begin{thebibliography}{99}

\bibitem{Girifalco}
L.~F.~Girifalco, J. Phys. Chem.  {\bf 95}, 5370 (1991); {\bf 96}, 858 (1992).
\bibitem{Spaepen} 
C.~M.~Lieber and C.-C.~Chen, {\it Solid State Physics} {\bf 48}, 109,
H.~Ehrenreich and F.~Spaepen eds. (Academic Press, San Diego, 1994).
\bibitem{Abramo1}
M.~C.~Abramo, C.~Caccamo, D.~Costa, and G.~Pellicane,
Europhys. Lett. {\bf 54}, 468 (2001).
\bibitem{Broughton}
J.~Q.~Broughton, J.~V.~Lill, and J.~K.~Johnson, 
Phys. Rev. B {\bf 55}, 2808 (1997).
\bibitem{Pacheco}
J.~M.~Pacheco and J.~P.~Prates-Ramalho, 
Phys. Rev. Lett. {\bf 79}, 3873 (1997);  
A.~L.~C.~Ferreira,  J.~M.~Pacheco and J.~P.~Prates-Ramalho,
J. Chem. Phys. {\bf 113}, 738 (2000).
\bibitem{Safran}
U.~S.~Schwarz and S.~A.~Safran, Phys. Rev. E {\bf 62}, 6957 (2000).
\bibitem{Hasegawa0}
O.~Umiguchi, T.~Inaoka, and M.~Hasegawa, 
J. Phys. Soc. Japan {\bf 68}, 508 (1999).
\bibitem{Schon}
J.~H.~Schon, C.~Kloc, and B.~Batlogg, Science {\bf 293}, 2432 (2001).
\bibitem{Noro}
M.~G.~Noro and D.~Frenkel, J. Chem. Phys. {\bf 113}, 2941 (2000).
\bibitem{short}
M.~H.~J.~Hagen and D. Frenkel, J. Chem. Phys. {\bf 101}, 4093 (1994); 
C.~F.~Tejero, A.~Daanoun, H.~N.~W.~Lekkerkerker, and M.~Baus,
Phys. Rev. Lett. {\bf 73}, 752 (1994).
\bibitem{Cheng}
A.~Cheng, M.~L.~Klein, and C.~Caccamo,
Phys. Rev. Lett. {\bf 71}, 1200 (1993).
\bibitem{Hasegawa}
M.~Hasegawa and K.~Ohno, J. Chem. Phys. {\bf 111}, 5955 (1999). 
\bibitem{Mooij}
M.~H.~J.~Hagen, E.~J.~Meijer, G.~C.~A.~M.~Mooij, D.~Frenkel,
and H.~N.~W.~Lekkerkerker Nature {\bf 365}, 425 (1993).
\bibitem{Hasegawa2}
M.~Hasegawa and K.~Ohno, J. Phys.:~Cond. Matter {\bf 9}, 3361 (1997).
\bibitem{Fucile}
C.~Caccamo, D.~Costa, and A.~Fucile, J. Chem Phys. {\bf 106}, 255 (1997).
\bibitem{Louis}
A.~A.~Louis, Philos.~T.~Roy.~Soc. A {\bf 359}, 939 (2001).
\bibitem{Chandler}
D.~Chandler, J.~D.~Weeks, and H.~C.~Andersen, Science {\bf 220}, 787 (1983).
\bibitem{Verlet}
J.-P.~Hansen and L.~Verlet, Phys. Rev. {\bf 184}, 151 (1969).
\bibitem{deltas}
P.~V.~Giaquinta and G.~Giunta, Physica A {\bf 187}, 145 (1992).
\bibitem{CMHNC}
C.~Caccamo, Phys. Rev. B {\bf 51}, 3387 (1995).
\bibitem{Coppolino}
M.~C.~Abramo and G.~Coppolino, Phys. Rev. B {\bf 58}, 2372 (1998).
\bibitem{Merida}
D.~Costa, C.~Caccamo, and M.~C.~Abramo, 
J. Phys.: Cond.~Matter {\bf 14}, 2181 (2002).
\bibitem{HM}
J.-P.~Hansen and I.~R.~McDonald, {\it Theory of Simple Liquids} 2nd ed.
(Academic Press, New York, 1986).
\bibitem{Ladd}
D.~Frenkel and A.~J.~C.~Ladd, J. Chem. Phys. {\bf 81}, 3188 (1984).
\bibitem{Smit}
D.~Frenkel and B.~Smit, {\it Understanding Molecular Simulation}
(Academic Press, London, 1996).
\bibitem{Widom}
B.~Widom, J. Chem. Phys. {\bf 39}, 2808 (1963).
\bibitem{ng} R.~E.~Nettleton and M.~S.~Green,
J. Chem. Phys. {\bf 29}, 1365 (1958).
\bibitem{spg} 
F.~Saija, S.~Prestipino, and P.~V.~Giaquinta,
J. Chem. Phys. {\bf 113}, 2806 (2000);
{\bf 115}, 7586 (2001), and references
therein. 
\bibitem{Cop2}
M.~Hasegawa and K.~Ohno, J. Chem. Phys. {\bf 113}, 4315 (2000).
\bibitem{WCA}
J.~D.~Weeks, D.~Chandler, and H.~C.~Andersen, 
J. Chem. Phys. {\bf 54}, 5237 (1971).
\bibitem{Ashcroft}
N.~W.~Ashcroft, Nature {\bf 365}, 387 (1993).
\bibitem{Foffi}
G.~Foffi \etal, Phys. Rev. E {\bf 65}, 031407 (2002); 
D.~Pini, J.~L.~Ge, A.~Parola, and L.~Reatto, Chem. Phys. Lett. 
{\bf 327}, 209 (2000).
\bibitem{Ballone}
D.~Costa, P.~Ballone, and C.~Caccamo, J. Chem. Phys. {\bf 116}, 3327 (2002). 
\bibitem{Pellican}
C.~Caccamo and G.~Pellicane, {\it to appear in} J. Chem. Phys. (2002).

\end{thebibliography}
\end{document}